# A Revised Collection of Sunspot Group Numbers

J.M. Vaquero[1,2] • L. Svalgaard[3] • V.M.S. Carrasco[2,4] • F. Clette[5] • L. Lefèvre[5] •

M.C. Gallego[2,4] • R. Arlt[6] • A.J.P. Aparicio[2,4] • J.-G. Richard[7] • R. Howe[8]

[1] Departamento de Física, Universidad de Extremadura, 06800 Mérida, Spain [e-mail: jvaquero@unex.es]

[2] Instituto Universitario de Investigación del Agua, Cambio Climático y Sostenibilidad (IACYS), Universidad de Extremadura, 06006 Badajoz, Spain

[3] W.W. Hansen Experimental Physics Laboratory, Stanford University, Stanford, CA 94305 USA

[4] Departamento de Física, Universidad de Extremadura, 06071 Badajoz, Spain

[5] World Data Center SILSO, Royal Observatory of Belgium, 3 Avenue Circulaire, 1180 Brussels

[6] Leibniz-Institut für Astrophysik Potsdam (AIP), An der Sternwarte 16, 14482 Potsdam, Germany

[7] Independent scholar, 6 rue Guesnault, 41100 Vendôme, France

[8] AAVSO, Solar Section, 49 Bay State Road, Cambridge, MA 02138, USA

**Abstract:** We describe a revised collection of the number of sunspot groups from 1610 to the present. This new collection is based on the work of Hoyt and Schatten (*Solar Phys.* **179**, 189, 1998). The main changes are the elimination of a considerable number of observations during the Maunder Minimum (hereafter, MM) and the inclusion of several long series of observations. Numerous minor changes are also described. Moreover, we have calculated the active-day percentage during the MM from this new collection as a reliable index of the solar activity. Thus, the level of solar activity obtained in this work is greater than the level obtained using the original Hoyt and Schatten data, although it remains compatible with a grand minimum of solar activity**.** The new collection is available in digital format.

## 1. Introduction

Telescopic observations of sunspots made since 1610 provide us an essential element to reconstruct the solar activity in the last four centuries (Vaquero and Vázquez, 2009; Clette *et al.*, 2014). The counting of sunspots has been described as the longest active experiment in the history of science (Owens, 2013). We need two essential elements for the reconstruction of solar activity: i) a collection as complete as possible of telescopic observations of sunspots and ii) a methodology to obtain a single time series from all records contained in the collection. Rudolf Wolf clearly understood the importance of i) in the sense that he made a monumental effort to obtain (and publish) the greatest possible number of historical records. Subsequently, Hoyt and Schatten (1998; hereafter HS98) conducted a new systematic survey in order to further increase the number of records, beyond what Wolf had already collected. However, the resulting extended data archive only included sunspot-group counts, as HS98 aimed at building a Group sunspot Number that did not include a count of individual sunspots.

The aim of this article is to describe a new, corrected version of the collection of sunspot-group counts based on the previous efforts by R. Wolf and HS98. In the last 15 years, several works have been published containing analyses of historical sources of sunspot observations (see references in Clette *et al.* (2014) for details). In this article, we document changes made to the HS98 data to obtain the revised collection of sunspot-group numbers. These changes include recently published additions/revisions to records in the HS98 data as well as modifications presented for the first time in this paper.

We should keep in mind that the compilation of group counts presented here only forms a first foundation for construction of a Group Sunspot Number (GSN) as a long-term measure of solar activity. Indeed, historical records can provide other elements about sunspots such as hemispheric asymmetry (Zolotova *et al.*, 2010), positions (Arlt *et al.*, 2013), areas (Vaquero, Sánchez-Bajo, and Gallego, 2002; Balmaceda *et al.*, 2009), or photospheric rotation rate (Casas, Vaquero, and Vázquez, 2006; Arlt and Fröhlich, 2012). An extensive use of historical sources related to sunspots should provide catalogs of sunspots (Arlt, 2009; Arlt *et al.*, 2013) including information not only about the number of sunspot groups. They could provide information about sunspot positions, areas, and even tilt angles of the sunspot groups (Senthamizh Pavai *et al.*, 2016). The scientific exploitation of these catalogs could be complex, because of a lack of common standards for the different sources of data (Lefèvre and Clette, 2014).

In this article, we provide information about the format and availability of our collection (Section 2), as well as a detailed description of changes and revisions for different time periods: early period (Section 3) and 19th – 20th centuries (Section 4). Additionally, we offer some conclusions and perspective for future work in Section 5.

**2. A Revised Collection**

The revised collection of sunspot group counts is contained in a machine-readable text file that is available at SILSO (sidc.be/silso/) and HASO ([haso.unex.es](haso.unex.es)). This file is divided into six columns. The first three columns contains the year, month, and day of the observation, respectively. The fourth column indicates the station number and the fifth column the observer of the station (both are zero if they are unavailable). Finally, the sixth column shows the number of sunspot groups (missing days are represented by the value -1). An example of the format is given in Table 1.

Table 1. Some example lines with the format of the data file.

| Year | Month | Day | Station | Observer | Groups |
|------|-------|-----|---------|----------|--------|
| 1610 | 1 | 1 | 0 | 0 | -1 |
| 1880 | 1 | 7 | 292 | 1 | 3 |
| 1880 | 1 | 7 | 318 | 1 | 2 |
| 1880 | 1 | 7 | 328 | 1 | 5 |
| 1880 | 1 | 7 | 332 | 1 | 3 |

Additionally, there is another file containing the list of sunspot observers. Each row gives the station number, the first and last year of observation, the total number of observations for that period, and the name of the observer (Table 2). Lastly, we have added a file describing the differences between this revised collection and the data provided by Hoyt and Schatten (1998) (Table 3).

Table 2. Some example lines of the list of sunspot observers.

| Station | Initial | Final | Tot. Obs. | Observer |
|---------|---------|-------|-----------|----------|
| 1 | 1610 | 1613 | 210 | HARRIOT, T., OXFORD |
| 2 | 1611 | 1640 | 882 | SCHEINER, C., ROME |
| 25 | 1642 | 1684 | 1656 | HEVELIUS, J., DANZIG |
| 332 | 1874 | 1976 | 37465 | ROYAL GREENWICH OBSERVATORY |

Table 3. Some example lines of the file containing the differences between this revised collection and Hoyt and Schatten (1998).

| Name | Year | Notes |
|---|---|---|
| ARGOLI, A., VENICE | 1634 | Added according to Vaquero (2003) |
| MARCGRAF, LEIDEN | 1637 | Added according to Vaquero et al. (2011) |
| CRABTREE, W., ENGLAND | 1639 | Estimated values. Removed |
| SIVERUS, H., HAMBURG | 1685 | Continuous values (zero). Removed |

This revised collection of sunspot-group counts contains more than one million observations by 738 different observers covering the period 1610 – 2010. Over these four centuries, temporal coverage is, of course, irregular. Figure 1 shows the number of days with records per decade in the revised collection presented in this article (dark-gray columns). Also shown is the corresponding temporal coverage for the HS98 database (light-gray columns). From 1610 to the start of Schwabe's observations in 1826, the number of days of observation per decade is lower in the revised collection than in the HS98 database because we discarded observations that we considered erroneous for various reasons.

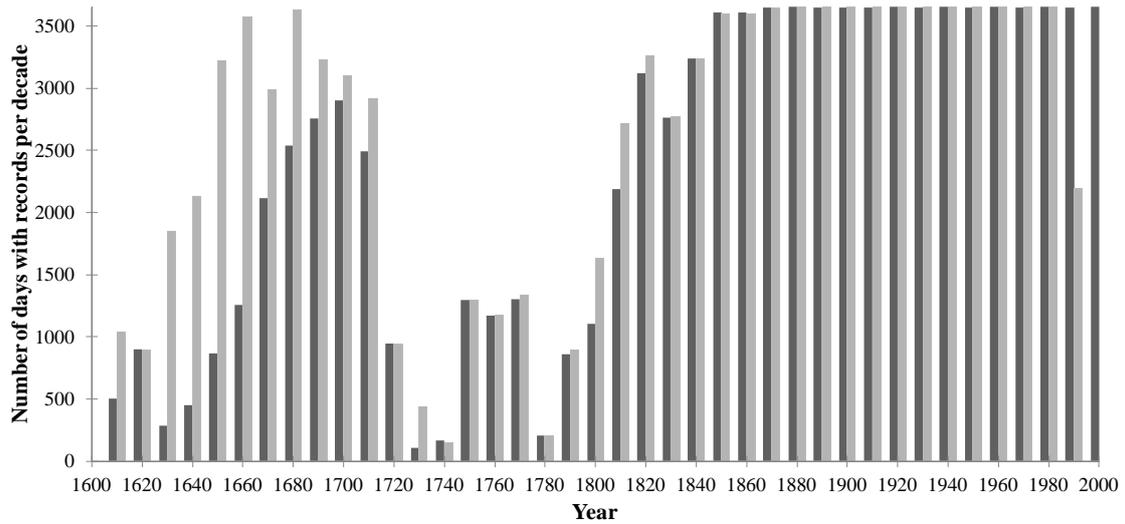

Figure 1. Number of days with records per decade in HS98 (light-gray columns) and in the revised collection presented in this article (dark-gray columns).

## 3. Revisions of Early Data

Our knowledge of solar activity in the historical era is derived from reconstructions from very sparse data. Therefore, it is important to obtain not only the largest possible number of observations, but also high-quality data. Recent articles have shown that HS98 included in their database a large number of counts derived from general mentions of the absence of sunspots on the solar disc and from astrometric measurements of the Sun such as solar-meridian observations (Clette *et al.*, 2014; Vaquero and Gallego, 2014). These kinds of data should be removed from the collection of the sunspot-group counts. Recent studies of explicit sunspot observations by Hevelius (Carrasco, Villalba Álvarez, and Vaquero, 2015b) and Flamsteed (Carrasco and Vaquero, 2016) have indicated that the general level of solar activity computed from explicit observations is significantly higher than that computed from general comments and astrometric records. Therefore, we have removed large parts of the HS98 database.

We also made an effort to re-count sunspot groups from original sunspot drawings. Thus, the sunspot drawings by Galileo, Gassendi, Staudacher, Schwabe, Wolf (small telescope), and Koyama were analyzed in order to obtain the number of sunspot groups using modern criteria based on the morphological classification of sunspot groups.

Moreover, we have incorporated in this revised collection some original observations that were not used by HS98. The case of Pehr Wargentin in 1747 is valuable because of the scarcity of records around that year. However, the main changes concern three different periods: i) the first years of observations (1610 – 1644), ii) the Maunder Minimum (MM) (1645 – 1715), and iii) the years around the "lost solar cycle" (1791 – 1797). The very recent data presented by Usoskin *et a*l. (2015) and Neuhäuser *et al*. (2015) have been also incorporated.

**3.1. The Earliest Years (1610 – 1644)**

The earliest years of our collection of observations, from 1610 to the beginning of the MM in 1645, are characterized by a great variability in the number of available counts. Generally, the number of observations per year is small.

We have added sunspot-group counts (not used until now) made by four early scientists: Argoli (Vaquero, 2003), Marcgraf (Vaquero *et al*., 2011), Strazyc (Vaquero and Trigo, 2014), and Horrox (Vaquero *et al*., 2011). Moreover, we have removed the observations attributed to Marius, Riccioli, and Zahn for the periods 1617 – 1618, 1618, and 1632 respectively, because they are an almost continuous list of zero-spot reports.

Finally, we have made two modifications in the counts by Horrox and Rheita according to the recent contributions by Vaquero *et al*. (2011) and Gómez and Vaquero (2015).

**3.2. The Maunder Minimum (1645 – 1715)**

Very recently, several studies fed a controversy about the true nature of MM from the phenomenological point of view (Zolotova and Ponyavin, 2015; Vaquero and Trigo, 2015; Vaquero *et al*., 2015; Usoskin *et al*., 2015). An important conclusion is that there is no doubt that at least some of the instruments used for solar observations during the MM were good enough to make an accurate count of sunspot groups. As an example, we can see in Figure 2 the equipment used by Hevelius. Therefore, a correct collection of the number of sunspot groups observed during the MM is crucial for further studies. In this section, we briefly describe the actions taken to obtain the revised collection. Basically, these actions can be split into three categories: i) the elimination of incorrect records, ii) the addition of newly found observations, and iii) the correction of counting errors.

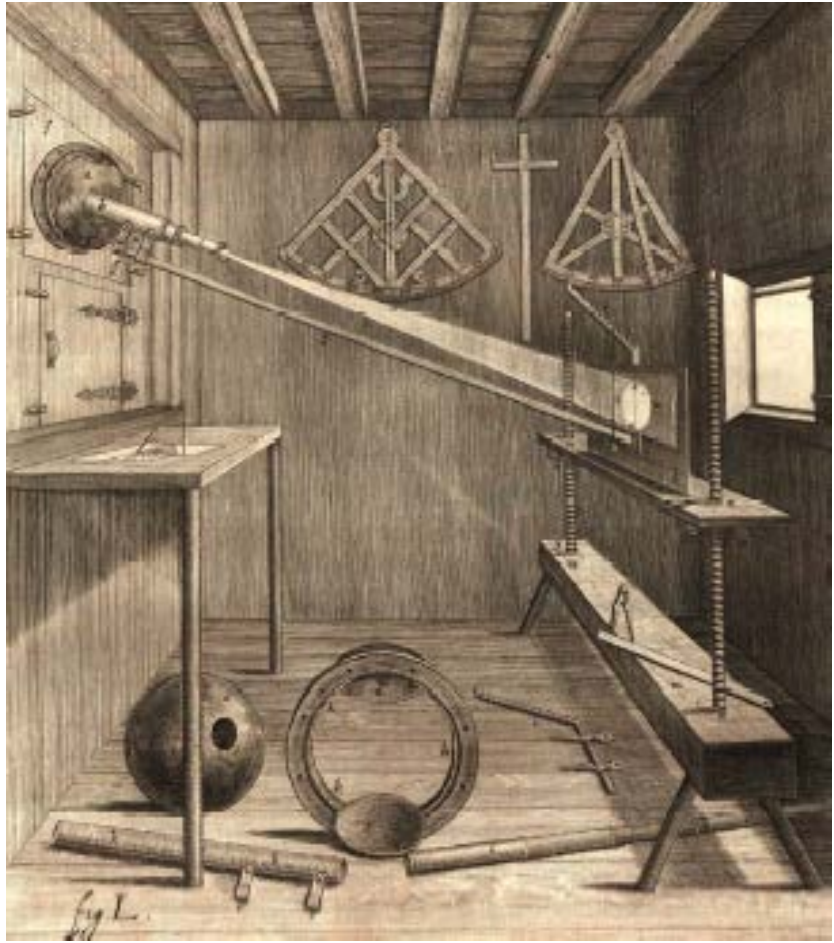

Figure 2. Astronomical instruments for solar observations used by Hevelius from his book *Machina Coelestis* (1679) (Courtesy of the Library of the Astronomical Observatory of the Spanish Navy).

We have discarded a large number of observations that were in the HS98 database during the MM, as stated above. Vaquero and Gallego (2014) indicated that records of sunspots made from astrometric observations should not be used for studies of solar activity in the past and may have a significant impact on the reconstructions of solar activity based on them. The most prominent example is formed by observations with the giant *camera obscura* of the Basilica of San Petronio in Bologna. Therefore, we have discarded the observations made with this instrument, which were included in the HS98 database. They are listed in Table 3 of Vaquero (2007). Moreover, observers listed in the HS98 database with ≈365 days of observations per year have been removed from the revised collection because these values (usually zero values) are based on general indirect comments and not on well-documented observations (see Section 3.2 of Clette *et al*., 2014). Finally, consulting the original documents, Usoskin *et al*. (2015) concluded that the sunspot observation assigned to Kircher in 1667 is erroneous and needs to be removed from the HS98 database. Therefore, we have discarded this record. Very few records of the MM have been added since the publication of the HS98 database. In our revised collection, we have now included the sunspot records by Peter Becker from Rostock (Neuhaeuser *et al*., 2015) and Nicholas Bion from Paris (Casas, Vaquero, and Vázquez, 2006).

In recent years, only one important change has been made in the counting during the MM. Vaquero, Trigo and Gallego (2012) used a simple method (based on the relationship between annual Group Sunspot Number and active days) to detect inconsistent values of the annual sunspot number in several years, including 1652. Later, Vaquero and Trigo (2014) detected that the origin of this problem is a misinterpretation of a comment by Hevelius describing his sunspot observation of 1652. Therefore, we have changed these observations in the revised collection.

The main modifications with respect to HS98 data are localized in this period. In terms of solar-activity level, we have also found noticeable differences between this work and HS98 during the MM. Figure 3 shows the statistics of the active-day percentage extracted from both articles for the period 1660 – 1712. The level of solar activity calculated from this revised collection (12,334 observation days with 8.8 % of active days) is greater than the level obtained by HS98 (17,557 observation days with 6.1 % of active days). However, including the beginning of the Maunder Minimum in the analysis (1640 – 1712), the percentage of active days from HS98 (22,915 observations) is 5.9 % while that the same percentage is 9.9 % from our revised collection (13,651 observations). In other words, the percentage of active days is almost doubled in this work compared to HS98.

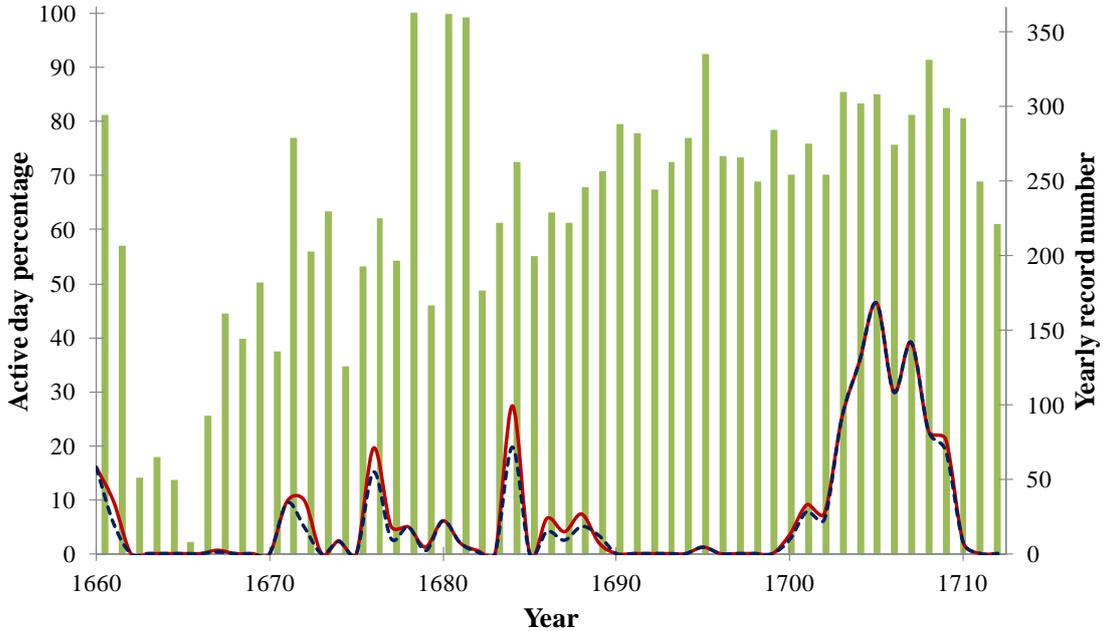

Figure 3. Statistics of the active-day percentage during the MM for HS98 (dashed-blue line) and this work (red line). The green bars represent the total number of yearly observations in the revised collection.

Although the level of solar activity calculated from this revised collection is greater than that calculated using HS98, this new result remains compatible with a grand minimum epoch of solar activity. From a sample of $n$ observation with $r$ active days, the most probable value of the fraction of active days in a year is given by the hypergeometric probability distribution (Kovaltsov, Usoskin, and Mursula, 2004). Thus, we have estimated the most probable value of the fraction of active days for the MM and the Dalton Minimum (Table 4). We find that the expected value for the MM is significantly less than for the Dalton Minimum, which is an epoch of reduced solar activity, although it is not considered to be a grand minimum. Therefore, the level of solar activity estimated from this revised collection confirms that the MM is a grand minimum.

Table 4. Expected value and upper and lower limits (99 % confidence interval) of the fraction of active days [%] estimated for the Maunder Minimum (1640 – 1712) and Dalton Minimum (1798 – 1833) from this revised collection.

| Period | Expected Value | Upper Limit | Lower Limit |
|---|---|---|---|
| Maunder Minimum | 9.94 % | 10.33 % | 9.55 % |
| Dalton Minimum | 61.63 % | 62.48 % | 60.78 % |

**3.3. Around the "Lost" Solar Cycle (1791 – 1797)**

A controversy about the presence of a "lost solar cycle" between the classical Solar Cycle 4 and 5 has divided the community over recent years (Usoskin *et al.*, 2009; Zolotova and Ponyavin, 2011; Owens *et al.*, 2015). Therefore, we have revised some sunspot records related to this controversial period.

We have analyzed the sunspot observations made by several astronomers (D. Huber, J.E. Bode, H. Flaugergues, F. von Hahn, F.A. von Ende, and J. Schröter) who were active during Solar Cycle 4. The aim of this analysis is to clarify and provide new information on this controversial "lost cycle".

We have revised the observations made by D. Huber. Note that observations made by D. Huber were improperly attributed to his father, Johann Jakob Huber, in the original HS98 database. In the latter, there is one record by Huber counting four sunspot groups on 28 May 1793. This is a very important record because observations around 1793 are very scarce. We have located the original document (Huber, 1808. *Brouillon für astron. Beob. 1793-1808*, p. 47) at the Library of the University of Basel: it is reproduced in

Figure 4a. We have changed the count from four to six sunspot groups. D. Huber noted in German: "My father had asked me to make this check, because a few hours ago Venus was (about to be) in conjunction with the Sun […]. He also recognized that they were clearly sunspots."

We have also revised the observations by J.E. Bode. We have modified the sunspot count for 3 April 1791 from five to six sunspot groups. The original sketch made by Bode (*Notebooks*, vol. 6, pp. 24 – 25) is reproduced in Figure 4b. Moreover, we have incorporated one additional record for 20 May 1794, when Bode recorded three sunspot groups (*Notebooks*, vol. 9, pp. 23 – 24). These manuscripts were located at the Archive of the Academy of Sciences of Berlin-Brandenburg.

H. Flaugergues was an important sunspot observer in this same time interval. His observations corresponding to the years 1794 and 1795 were made in Aubenas (not in Viviers). We have removed the records assigned to "H. Flaugergues (C. de T.)" in the HS98 database because they include continuous null spot counts and show inconsistences with the observations reported by the same observer (H. Flaugergues) from Viviers. We have also corrected a total of 17 records using the original documents (one record in 1788, four in 1794, seven in 1795, and five in 1796). We have lowered the sunspot counts in 13 observations and have increased it in four other cases. Moreover, another 34 sunspot counts have been added to the revised collection (one in 1795, 9 in 1796, and 24 in 1797). These sunspot-group counts from new observations range from zero to two. Two original sources have been consulted: i) the manuscripts located at Library of the Paris Observatory (Flaugergues, *Astronomie du 12 Novembre 1782 au 21 Septembre 1798*) and ii) the records contained in the journal *Mémoires de l'Institut National des Sciences et Arts*.

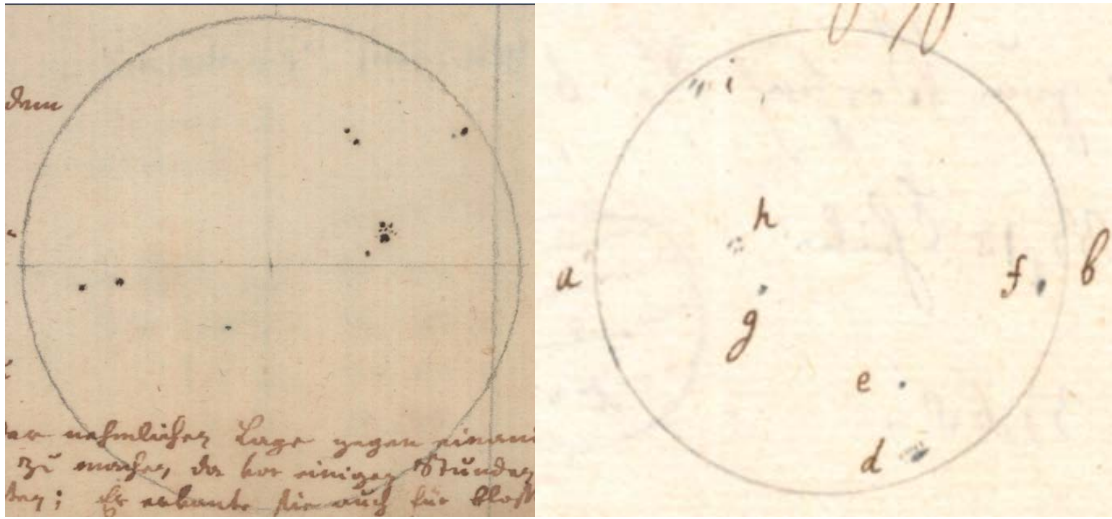

(a) (b)

Figure 4. Original sketches of sunspot observations by (a) Huber (28 May 1793) and (b) Bode (3 April 1791) [Sources: (a) Huber, 1808. *Brouillon für astron. Beob. 1793 – 1808*, Sign. L lb 12, fol. 47, Library of University of Basel, and (b) Bode, 1791, *Notebooks*, vol. 6, pp. 24 – 25, Archive of the Academy of Sciences of Berlin-Brandenburg].

We have also revised the records by F. von Hahn, incorporating a lost record (4 February 1793) when no sunspot group was observed. This record can be consulted (in German) in *Berliner Astronomisches Jahrbuch* ("Remarks about Venus, Description of some remarkable sunspots, and astronomical news. Submitted from May 13th to June 16th 1793", pp. 188 – 191, Berlin, 1793). Moreover, we did not modify data from F.A. von Ende that ha**s** also been reviewed.

Finally, three observations by J. Schröter in the year 1795 (30 November, 3 and 5 December), when he recorded one sunspot group in each of the three cases, were added to the collection. These observations (in German) lie in *Neuere Beyträge zur Erweiterung der Sternkunde* (Chapter VI: Observation of a remarkable and astonishing sunspot, together with further remarks about the natural constitution of the Sun (Lilienthal, 1 February1796), pp. 56 – 77, Göttingen, 1798).

According to this new revision of sunspot observations in the 1790s, the "lost solar cycle" seems less plausible due to the confirmation of a relatively high number of sunspot groups in 1792 and 1793. However, we stress that a more detailed study is necessary incorporating more information, namely the heliographic latitudes of the sunspots.

## 4. New Series in the 19th and 20th Centuries

### 4.1. D.E. Hadden

D.E. Hadden made sunspot observations in Alta (Iowa, USA) during the period 1890 – 1931. However, we were able to recover only 13 years from the 42 years of observations with daily information. In total, 2964 daily counts have been recovered. These counts were published in several astronomical journals; some of them were local journals. Moreover, Hadden used different telescopes in each observation period: i) 1890 – 1896, three-inch (≈76 mm) refractor telescope and ii) 1897 – 1902, four-inch (≈100 mm) refractor telescope.

In their collection, HS98 included records by Hadden only for the last third of 1890 (67 observations in total). However, these values are incorrect since they only report new groups that appeared on the solar disk and not the total number of groups present on the Sun (Carrasco *et al.*, 2013).

### 4.2. Madrid Observatory

The Astronomical Observatory of Madrid (AOM) was founded in the late 18th century. Systematic observations were made from 1876 to 1986 to determine the sunspot number and area. The data were published in various Spanish scientific publications. Aparicio *et al.* (2014) retrieved and digitized these data. From the group and sunspot counts, they computed the Madrid sunspot number (MSN) and the Madrid group sunspot number (MGSN). The subsequent analysis showed that the MSN and the MGSN can be considered as reliable series given their very high correlation with other reference indices.

In addition, Aparicio *et al.* (2014) recovered interesting metadata about the instruments, methods, and observers of the AOM solar program. These metadata reveal some mistakes in the construction of the Group Sunspot Number (GSN) by HS98. They considered Aguilar to be the observer for the period 1876 – 1882 and Merino for 1883 – 1896. However, the observer for those years was Ventosa. Aguilar and Merino acted only as directors of the observatory in those respective periods. Later, HS98 took observations for the years 1935 – 1938, 1940 – 1957, and 1959 – 1972 with "Madrid Observatory" as the observer name. However, as has been shown by Aparicio *et al.* (2014), two important facts must be taken into account. Firstly, the observations for the years 1937 – 1938 were made in Valencia (due to the Spanish Civil War) by other observers with other instruments. Secondly, in the period 1935 – 1986, there were a large number of observers with an irregular distribution. For those reasons, we must be very careful when working with these data in order to calculate the correction factors. Lastly, Aparicio *et al.* (2014) added a large quantity of available sunspot-group data that were not used by HS98. Thus, the daily observations from AOM corresponding to the periods 1876 – 1896 and 1936 – 1986 have been included in this revised collection. We emphasize that the observations made at AOM from 1973 to 1986 are not included in HS98 database. The total number of these new records is equal to 2936.

**4.3. Harry B. Rumrill**

Harry Barlow Rumrill followed closely in the footsteps of his friend Alden Walker Quimby in observation of sunspots. Rumrill's estate included an archive of his sunspot work. The archive is now in the possession of John Koester, of the Antique Telescope Society (New York, NY). The**se** raw data consist of a large collection of pencil drawings of the solar disk. A few thousand drawings dating from 8 January 1928 through 31 October 1950 (with gaps) are present. Raw data from the drawings were summarized in smaller notebooks, which were photocopied by Koester and forwarded to one of us [L. Svalgaard]. Each page in these notebooks gives data in six columns, one of which is subdivided into two parts. They are labeled: date; time; new groups; total groups/total spots (these given in one subdivided column under the heading "Total"); groups faculae (*sic*); definition (in addition to a description of observing conditions, this column often has a note as to which telescope was used) (see Figure 5).

![Sun Spots and Faculae notebook page]

| 1939 | Time | New Groups | TOTAL Groups | Spots | Groups Faculae | Definition |
|---|---|---|---|---|---|---|
| June 27 | 8:30 a | — | 6 | 33 | 2 | Fair |
| | | Much cloud | | | | |
| 28 | 8:30 a | 1 | 6 | 37 | 4 | Poor |
| | | Much cloud. New group near eastern limb. | | | | |
| 30 | 10 a | — | 4 | 32 | 4 | Good |
| July 1 | 7:30 a | 1 | 5 | 39 | 3 | Good |
| | | Probably should be reckoned as 6 groups. | | | | |
| 2 | 7:30 a | — | 6 | 35 | 5 | Good |
| 3 | 7:15 a | 4 | 9 | 23 | 5 | Good |
| 4 | 8:45 a | — | 9 | 36 | 5 | Beautiful |
| | | The faculae especially fine. | | | | |
| 6 | 7:30 a | 2 | 6 | 51 | 2 | Good |
| | | Much cloud | | | | |
| 8 | 7:30 a | 1 | 5 | 77+ | 2 | Good |
| | | A magnificent exhibition | | | | |

Figure 5. An example page from the Rumrill notebooks (courtesy of John Koester).

**4.4. Herbert Luft**

Herbert Luft has one of the longest series of sunspot observations of this revised collection. His observations begin in 1923 and end in 1987. The series is thus 65 years long, although there are some gaps concentrated in the 1930s and 1960s. Luft made his observations in several parts of the world, first in Germany, then Brazil, then the US. He was detained at the Buchenwald concentration camp for five weeks in 1938 and, therefore, decided to immigrate to Brazil in 1939. His refractor telescope of 52 mm diameter was one of the few personal effects that accompanied him on this trip. He belonged to several astronomical associations and was mentored by the German astronomer Wolfgang Gleissberg, who recommended Luft focus on the observation of sunspots (Mattei and Mattei, 1989).

The original database of HS98 contains no records by H. Luft. Of the nearly 12,000 pages of sunspot observations in the notebooks of Luft, one of us [L. Svalgaard] has recovered those with good image quality. Thus, 10,628 new daily records made by Luft are now incorporated into the revised collection.

### 4.5. Thomas A. Cragg

Thomas A. Cragg joined the AAVSO in 1945 at age 17, when he was working as an assistant at the Mt. Wilson Observatory in California (Figure 6, left). He made an impressive total of over 157,000 variable-star observations (AAVSO Observer Initials CR), but he was equally dedicated to his daily solar observing (AAVSO Solar Initials CR), which spanned the years 1947 through 2010. Each sunspot count recorded in his observing journal included a drawing of the group and spot configurations (Figure 6, right).

Cragg lived in Los Angeles until he was about 48 (thus, around 1976). Then he moved to Australia and worked at the Siding Spring Observatory, as well as continuing his observing. After his death in 2011, his wife Mary sent all of his solar (and variable star) records to AAVSO Headquarters for the AAVSO archives. Mike Saladyga and Sara Beck have entered these solar data into the SunEntry solar database at AAVSO Headquarters. These data have been included in our revised collection of Sunspot Group Numbers.

In 1947, the AAVSO began collecting sunspot data from 23 observers, including Cragg, all of whom contributed to the American Relative Sunspot Number Index [$R_a$] generated using the data submitted to the AAVSO. This was the start of the AAVSO's Solar Division (now Solar Section). At that time (and until recently), the paper report forms containing observers' raw data were not saved once the $R_a$-number had been generated. Without the paper forms, and with no way to save the**se** data electronically**,** for many years the AAVSO historical solar raw data were lost.

Recovery of original sunspot data is possible, however, when observers' solar observing notebooks are made available for digitization. Recently, longtime solar observer Herbert Luft's nearly 70 years' of sunspot data were recovered from his notebooks at the AAVSO (see Section 4.4). Now, we have the Thomas Cragg drawings digitized in the SunEntry database as a continuous record of group, sunspot counts, and Wolf numbers. Thus, 17,726 new daily records made by Thomas Cragg are now incorporated into the revised collection.

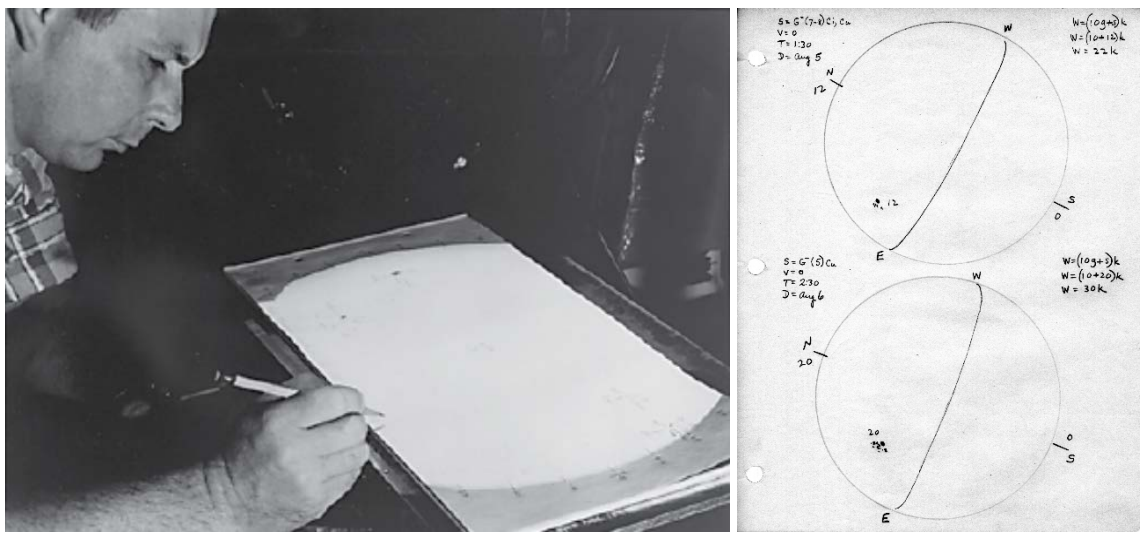

Figure 6. AAVSO member and observer Tom Cragg at work in the Mt. Wilson Solar Observatory (September 1962) (left) and an example page of his notebooks (right).

**4.6. Astronomical Observatory of the University of Valencia**

The Astronomical Observatory of the University of Valencia (Spain) was founded in 1909 by Ignacio Tarazona y Blanch. This observatory developed a solar-monitoring program, with astronomer Tomas Almer Arnau responsible for the observations. Sunspot counts were based on photographic plates. The Observatory published a catalog of sunspots for the period 1920 – 1928, except for 1921 – 1922 (Carrasco *et al*., 2014). Furthermore, it had good equipment, in particular a refractor telescope by Grubb with a 152 mm aperture. These observations made at the Observatory of Valencia were not compiled by HS98 in their database. Therefore, we have incorporated a total 1893 days with new observations in the revised collection, representing approximately 74 % of all days over the studied period.

**4.7. Data from the World Data Center SILSO**

For the most recent part of the database after 1980, we imported the group counts from the extensive sunspot-number database of the World Data Center SILSO (Clette *et al*. 2007, 2014). This database includes all observations collected on a monthly basis from the worldwide network coordinated by the WDC–SILSO, for a current total of more than 530,000 individual daily observations. In total, 282 stations contributed since 1980, with on average 85 stations active at any given time and between 20 and 45 observations available on any given day. Among all stations, 80 long-duration stations provided data over more than 11 years, some for more than 35 years. Two-thirds of the observers are individual amateur astronomers and one-third of the stations are professional observatories, often with different observers serving at different times.

Given the abundance of observations, this part of the database allows extensive statistical diagnostics for the determination of the Group sunspot Number.

When importing group data from the SILSO database, we used the standard two-letter station identifier, as defined for all SILSO stations since 1980 and still currently in operational use.

**5. Conclusion and Future Work**

We have presented a revised collection of sunspot-group numbers. Our collection has a smaller number of observations than the HS98 database for the early historical period. According to the records corresponding to the 17th and 18th centuries, the HS98 database contains 58,615 observations and 35,045 observation days while this new collection has 31,480 observations and 23,120 observation days. Conversely, for the period 1800 – 2010, the new revised collection (1,020,934 observations and 71,143 observation days) has a larger number of records than HS98 database (396,627 observations and 66,844 observation days). Thus, for the entire period 1610 – 2010, the new revised collection has 1,052,414 total observations with 94,263 observation days while the HS98 database contains 455,242 total observations with 101,889 observations days. Moreover, the quality of observations has been much improved, many typographical errors have been fixed, and an update has been made. Thereby, we have incorporated to the new revised collection more than 500,000 observations (approximately 1,000 new or corrected records correspond to the period 1610 – 1799) and more than 30,000 observations (about 25,000 records for the 17th and 18th centuries) have been discarded with respect to the HS98 database.

A large number of observations that we have discarded are reports of a spotless Sun during the MM. These records were associated with astrometric observations of the Sun. In fact, some of these observations were made using pinhole cameras (not telescopic devices). Furthermore, we have calculated the statistics of active days during the MM. We emphasize that although the level of solar activity extracted from this new collection is greater than the level obtained from HS98, our new results confirm that the MM is a grand minimum of solar activity. Thus, this contradicts the work of Zolotova and Ponyavin (2015).

The experience acquired during the process of compiling this collection has shown that records of sunspot groups can still be improved. The recent addition of supposedly lost observations, such as the observations by Marcgraf (Vaquero *et al*., 2011), Wargentin, or Peter Becker (Neuhäuser *et al.*, 2015), illustrates how a meticulous inquiry in historical archives and libraries could still offer surprising data for our international community. Moreover, the language used in the historical reports is mainly Latin. Thus, the translation from Latin to a modern language such as English is a priority task and some efforts have been made recently (Carrasco, Villalba Álvarez, and Vaquero, 2015a; Carrasco, Villalba Álvarez, and Vaquero, 2015b; Gómez and Vaquero, 2015). Therefore, we hope to update the revised collection of sunspot groups presented in this article every two or three years, publishing the updated files in several web sites including SILSO (sidc.be/silso/) and HASO (haso.unex.es).

Nevertheless, this remains an ongoing work. The possibilities offered by historical observations are vast, and so far we only derived the number of sunspot groups from them. An immediate first step should be to complement this collection with the total number of spots in each observation. The second step should be the compilation of hemispheric values. Both tasks would give us useful data to undertake further studies of the evolution of solar activity during the last four centuries.

**Acknowledgements**

J.-G. Richard acknowledges the help of Dr. Heiligensetzer (head of Historical Archives at the Library of the University of Basel) for his help in reading the manuscript of Daniel Huber, Vera Enke (Head of the Archives of the Academy of Sciences of Berlin-Brandenburg) for her help in reading the notebooks of J.E. Bode, and Sandrine Marchal (Head of the "Fonds Ancien") and Josette Alexandre at the library of the Paris Observatory for their help in reading Honoré de Flaugergues' notebooks. We thank the referee for the helpful comments. The authors have been benefited from the participation in the Sunspot Number Workshops. L. Svalgaard thanks the AAVSO for their support in recording the notebooks by Luft. A.J.P. Aparicio thanks the Ministerio de Educación, Cultura y Deporte for the award of a FPU grant. F. Clette and L. Lefèvre would like to acknowledge financial support from the Belgian Solar-Terrestrial Center of Excellence (STCE: www.stce.be). This work was partly funded by FEDER-Junta de Extremadura (Research Group Grant GR15137) and from the Ministerio de Economía y Competitividad of the Spanish Government (AYA2014-57556-P).

**Disclosure of Potential Conflicts of Interest**

The authors declare that they have no conflicts of interest.